\documentclass[journal=apchd5,manuscript=letter]{achemso}

\usepackage[version=3]{mhchem} % Formula subscripts using \ce{}
\usepackage[T1]{fontenc}       % Use modern font encodings

\usepackage[usenames, dvipsnames]{color}

\author{Christopher R. Gubbin}
\author{Simone De Liberato}
\email{S.De-Liberato@soton.ac.uk}
\affiliation{School of Physics and Astronomy, University of Southampton, Southampton, SO17 1BJ, United Kingdom}

\title{Theory of four-wave-mixing in phonon polaritons}

\begin{document}

\begin{abstract}
Third order anharmonic scattering in light-matter systems can drive a wide variety of practical and physically interesting processes from lasing to polariton condensation.
Motivated by recent experimental results in the nonlinear optics of localised phonon polaritons, 
in this Letter we develop a quantum theory capable of describing four-wave mixing  in arbitrarily inhomogeneous photonic environments. Using it we investigate Kerr self-interaction and parametric scattering of surface and localised phonon polaritons, showing both processes to be within experimental reach.
\end{abstract}

Polaritons are quasiparticle excitations arising from strong coupling between light and matter. They inherit the best properties of both constituents, being light and fast but also strongly interacting and therefore present an ideal platform for nonlinear photonics. In microcavity polariton systems \cite{Houdre94}, four-wave-mixing processes mediated by the Coulomb interaction between underlying excitonic degrees of freedom allow for realization of polariton condensation \cite{Kasprzak06} and superfluidity \cite{Amo09} as well as of practical devices such as polariton parametric oscillators \cite{Baumberg00,Diederichs06} and electrically pumped polariton lasers \cite{Bhattacharya13}.\\
When photons hybridise with coherent charge oscillations close to an interface, the resulting evanescent polariton modes can be localised on sub-wavelength scales. Plasmon polaritons at the surface of noble metals are archetypical, but their lossy character due to strong intrinsic Ohmic heating, have until now limited their practical applications \cite{Khurgin15}.
Phonon polaritons whose matter component is due to the ions of a polar dielectric, represent in this sense an interesting low loss option, covering the mid-infrared spectral region \cite{Caldwell15}. Not only have they been exploited for coherent thermal emission \cite{Schuller09}, but more recently tunable, localized resonances have been observed in Silicon Carbide (SiC) microresonators \cite{Caldwell13, Chen14, Gubbin17a,Wang17}, with quality factors  far exceeding the theoretical limit for plasmon polaritons. Localized phonon polaritons therefore present an ideal platform to translate established plasmonic and polaritonic technology to the mid-infrared. 
To quantify the potential of surface phonon polaritons for mid-infrared nonlinear and quantum optics we recently developed a real space Hopfield-like framework which allows for quantization and diagonalization of arbitrary inhomogeneous systems in terms of polaritonic excitations whose field profile can be obtained by solving inhomogeneous Maxwell equations \cite{Gubbin16b}.  We initially used such a theory to describe nascent second harmonic generation experiments on flat SiC surfaces \cite{Gubbin17b}. Those nonlinear processes have now been demonstrated experimentally utilising thick SiC substrates in reflectance \cite{Paarmann16}, from sub-diffraction surface phonon polaritons excited by prism coupling \cite{Passler17}, and from arrays of discrete SiC nano-resonators \cite{Razdolski16}.\\ 
In this Letter we develop the theory of phonon polariton four-wave-mixing, demonstrating that two important applications are within experimental reach with available sources. The first, Kerr self-interaction \cite{DeLeon14}, would be the first experimental verification of third order nonlinearities in SiC, allowing us to experimentally fix the material scattering parameters beyond current best estimates based on {\it ab initio} simulation \cite{Vanderbilt86}. The second, polaritonic scattering in a parametric oscillator configuration, will demonstrate the viability of SiC phonon polariton resonators for mid-infrared quantum polaritonics and nonlinear optics. \\
Following the procedure previously demonstrated \cite{Gubbin16b} and detailed in the Supplementary Information we can write the general third-order nonlinear Hamiltonian as
\begin{equation}
	\hat{\mathcal{H}}^{(3)} = \frac{3}{4} \epsilon_0 \sum_{ijkl} \sum_{\left[\mathbf{p}_1-\mathbf{p}_4\right]} \int \mathrm{d}\mathbf{r} \;  \chi_{ijkl}^{(3)} \left(\mathbf{r},\omega_{\mathbf{p}_1},\omega_{\mathbf{p}_2},\omega_{\mathbf{p}_3},\omega_{\mathbf{p}_4}\right)  \hat{\mathrm{E}}_{\mathbf{p}_1}^{i}\left(\mathbf{r}\right) \hat{\mathrm{E}}_{\mathbf{p}_2}^{j}\left(\mathbf{r}\right)\hat{\mathrm{E}}_{\mathbf{p}_3}^{k}\left(\mathbf{r}\right)\hat{\mathrm{E}}_{\mathbf{p}_4}^{l}\left(\mathbf{r}\right), \label{eq:HamNL}
\end{equation}
where vector components are denoted by $i$,$j$,$k$,$l$.  We can write the electric field of each mode as a function of the polaritonic operators $\hat{\mathcal{K}}_{\mathbf{p}}$
\begin{equation}
	\hat{\mathbf{E}}_{\mathbf{p}}\left(\mathbf{r}\right) = \sqrt{\frac{\hbar \omega_{\mathbf{p}}}{2 \epsilon_0 \mathrm{V}_{\mathbf{p}}}} \left(\bar{\boldsymbol{\alpha}}_{\mathbf{p}}\left(\mathbf{r}\right) \hat{\mathcal{K}}_{\mathbf{p}} + \boldsymbol{\alpha}_{\mathbf{p}} \left(\mathbf{r}\right)\hat{\mathcal{K}}_{\mathbf{p}}^{\dag} \right),
\end{equation} 
where $\mathrm{V_{\mathbf{p}}}$ the is the photonic mode volume \cite{Gubbin16b},  $\boldsymbol{\alpha}_{\mathbf{p}}\left(\mathbf{r}\right)$ are the Hopfield coefficients obtained solving the Maxwell equations in the inhomogeneous system, and $\mathbf{p}$ is a composite index over the polaritonic normal modes.

\section*{Results and Discussion}
In this Letter we will consider the zincblende polytype $\beta$-SiC, whose dielectric function is negative in the Reststrahlen band between the transverse and longitudinal optic phonon frequencies $\omega_{\mathrm{TO}} = 797.5/$cm and $\omega_{\mathrm{LO}}= 977/$cm. We consider the high-frequency dielectric constant $\epsilon_{\infty} = 6.49$ and transverse phonon damping rate $\gamma = 4$/cm\cite{Paarmann16}. In this region $\beta$-SiC supports surface phonon polaritons, allowing sub-diffraction light localisation by transient storage of energy in the kinetic and potential energy of the ionic lattice. 
Surface phonon polaritons are a promising platform for mid-infrared nonlinear optics due to their ultra-low mode volumes, high quality factors, and the intrinsic anharmonicity of the host polar dielectric. As we lack measurements of the required nonlinear parameters for $\beta$-SiC, in this Letter we utilise the {\it ab initio} results of Vanderbilt {\it et al.}, who studied second and third order anharmonic phonon-phonon scattering in various materials with a diamond crystal structure. They observed a certain universality in coefficients governing the mechanical anharmonicity due to phonon-phonon scattering on rescaling by the bond length and spring constant \cite{Vanderbilt86}. Our recent results demonstrate that such values lead to reasonable estimates, at least for second order nonlinearities, also for $\beta$-SiC  \cite{Gubbin17b}, and we expect the same should hold for third order nonlinearities. Intriguingly, the predicted third-order nonlinear anharmonic scattering coefficients are attractive, contrasting the repulsive material interactions in microcavity polariton systems. As shown in the Supplemental Information the $\chi^{(3)}$ nonlinear susceptibility is composed of a sum of resonant terms, with poles of order 1 to 4 close to the optical photon frequency $\omega_{\mathrm{TO}}$. Given that all modes participating in the processes studied in this Letter lie in the Reststrahlen band, in the neighbourhood of the highest fourth order pole of $\chi^{(3)}$, we can ignore lower order contributions to the optical nonlinearity and consider solely the pure mechanical contribution due to phonon-phonon scattering. Furthermore in $\beta$-SiC only the $xxxx, xyxy, xxyy$ components of the nonlinear tensor are non-zero when accounting for screening \cite{Vanderbilt86}.\\ 
In nonlinear optical experiments the available pump fluence is a key parameter determining which phenomena are within experimental reach. We therefore use parameters representative of the free-electron laser previously utilized to probe $\chi^{(2)}$ nonlinearities in polar dielectrics by the group of A. Paarmann \cite{Passler17}, specifically a pulsed excitation of duration $1$pS, energy $2\mu$J, and an elliptical excitation spot with semi-major and semi-minor axes $500\mathrm{\mu}$m and $150\mathrm{\mu}$m respectively corresponding to a pulse fluence of $0.84$mJ/cm$^2$.

\subsection*{Kerr Self-Interaction}
\begin{figure}
	\includegraphics[width=0.35\textwidth]{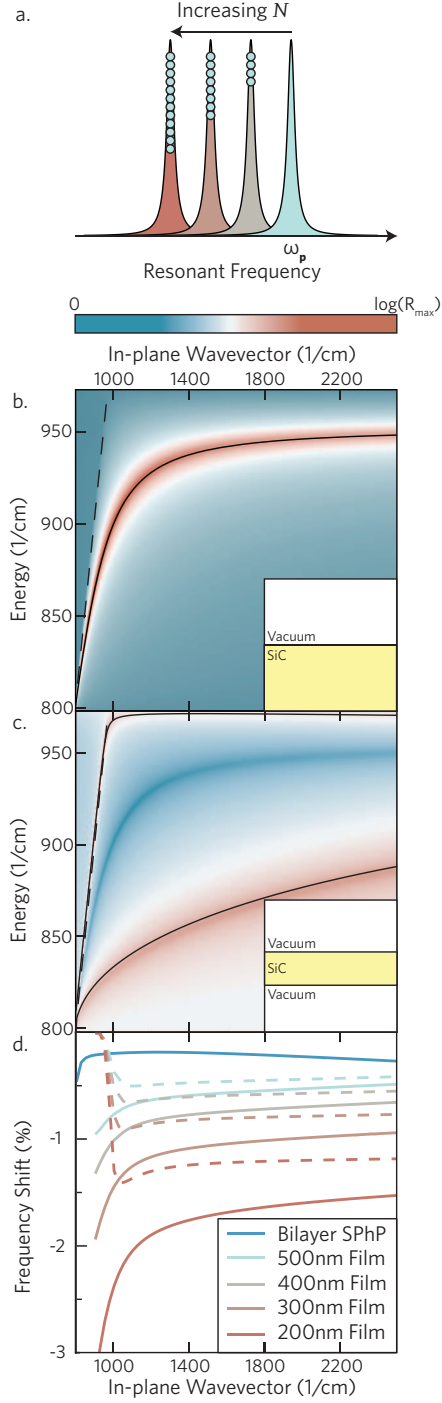}
	\caption{\label{fig:Fig1} a. Sketch of self-interaction, as a frequency shift dependent on excitation number $\mathcal{N}$ from bare frequency $\omega_{\mathbf{p}}$. b. Reflectance of the bilayer SiC/ vacuum interface illustrated in the inset. c. Reflectance of the vacuum/ $200$nm SiC/ vacuum trilayer shown in the inset. On plots b and c the solid line indicates the poles of the reflectance coefficient, signifying the polariton dispersion. Both colour maps are plotted on a logarithmic scale. d. The frequency shift calculated for the bilayer mode and the film modes (solid lines indicating the lower branch and dashed lines the upper) from Eq.~\ref{eq:SPMre} for pump parameters representative of a free electron laser \cite{Passler17}.}
\end{figure}
The Kerr effect manifests as an intensity dependant frequency shift due to the excitations self-interaction, interpretable as an intensity dependant shift in the effective refractive index \cite{DeLeon14b}. The nonlinear Hamiltonian in Eq.~\ref{eq:HamNL}, specialised onto the case of self-interaction ($\mathbf{p}_j = \mathbf{p}$, $\forall j$) and linearised, leads to renormalised modal frequencies $\omega_{\mathbf{p}}' = \omega_{\mathbf{p}}\left(1 + \Delta_{\mathbf{p}}\right)$ as sketched in Fig.~\ref{fig:Fig1}a. The fractional shift is derived in the Supplementary Information and given by
\begin{equation}
\label{eq:SPMre}
	\Delta_{\mathbf{p}} =  \frac{9}{2} \frac{\hbar \omega_{\mathbf{p}} \mathcal{N}_{\mathbf{p}}}{\epsilon_0 \mathrm{V}_{\mathbf{p}}^2}  \sum_{ijkl}  \int \mathrm{d}\mathbf{r} \;  \chi_{ijkl}^{(3)} \left(\mathbf{r}, \omega_{\mathbf{p}},\omega_{\mathbf{p}},\omega_{\mathbf{p}},\omega_{\mathbf{p}}\right)   \bar{\alpha}_{\mathbf{p}}^{i}\left(\mathbf{r}\right) \bar{\alpha}_{\mathbf{p}}^{j}\left(\mathbf{r}\right) \alpha_{\mathbf{p}}^k\left(\mathbf{r}\right) \alpha_{\mathbf{p}}^l\left(\mathbf{r}\right),
\end{equation}
with $\mathcal{N}_{\mathbf{p}}$ the population in the mode under consideration.
In order to assess whether such effects are large enough to be measured we initially consider two 1D inhomogeneous systems, consisting of either a planar $\beta$-SiC halfspace (inset of Fig.~\ref{fig:Fig1}b) or a thin, suspended, $\beta$-SiC film, such as those usually utilised for fabrication of planar photonic crystals \cite{Calusine14} (inset of  Fig.~\ref{fig:Fig1}c). The halfspace system supports a single evanescent surface phonon polariton while the film supports both symmetric and anti-symmetric superpositions of the surface phonons at each interface. The modes of the system, found considering the poles of the reflection coefficients for the stack, are shown in Figs.~\ref{fig:Fig1}b and \ref{fig:Fig1}c for the bilayer system and a suspended SiC film of thickness $200$nm. \\
Utilising Eq.~\ref{eq:SPMre} we may now calculate the expected fractional frequency shifts, shown in Fig.~\ref{fig:Fig1}d. Larger shifts are observed for thinner films, understandable as a result of increased in-plane confinement for the upper polariton branch and a depression in frequency toward the pole in material $\chi^{(3)}$ at $\omega_{\mathrm{TO}}$ for the lower branch. The decrease in shift for the upper branch at low in-plane wavevector is a result of the mode moving close to the lightline. These modes can be excited via prism coupling \cite{Passler17}. It was assumed that excitation occurred near critical coupling, where approximately $80\%$ of the input beam couples to the surface mode. \\
\begin{figure}
\includegraphics[width=0.35\textwidth]{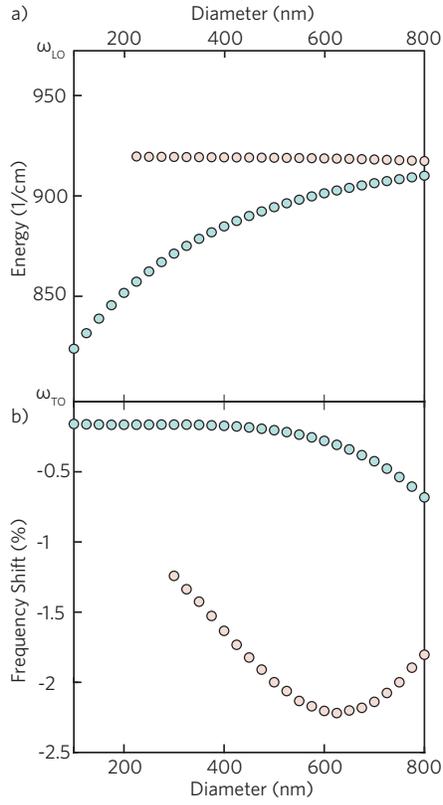}
\caption{\label{fig:Fig2} a. Real resonant frequencies of the lowest order transverse (m=1, red circles) and longitudinal (m=0, blue circles) modes of the single, cylindrical resonator shown in the inset as a function of diameter. The cylinder height is held at $800$nm and the corners are smoothed with radius $2$nm. b. Fractional frequencies shifts for the same cylinder calculated by Eq.~\ref{eq:SPMre}}	
\end{figure}
In order to further explore the nonlinear shift we now consider the modes of sub-wavelength resonators, which support optical resonances with ultra-small photonic mode volumes and correspondingly high energy localisation \cite{Chen14, Gubbin17a}. Here we simply study an isolated cylindrical resonator whose diameter is varied from $800$nm to $50$nm but our results should provide an accurate approximation to those expected for experimentally practical resonator arrays \cite{Gubbin16a}. The modes of such a resonator vary azimuthally as $e^{i m \phi}$ and they are easily probed utilising   COMSOL Multiphysics, allowing for evaluation of the integrated quantities entering Eq.~\ref{eq:SPMre}.\\
The modal frequencies for the lowest order longitudinal $(m=0)$ and transverse $(m=1)$ modes are shown in Fig.~\ref{fig:Fig2}a. The transverse mode is weakly affected by the geometrical change as the cylinder is quasistatic for all sizes studied, while the longitudinal mode shifts to the transverse optic phonon frequency as the diameter decreases. In Fig.~\ref{fig:Fig2}b we show the expected frequency shift. The shift for the transverse mode is far larger than for the longitudinal mode despite the longitudinal mode lying closer to the pole in the $\chi^{(3)}$ at $\omega_{\mathrm{TO}}$ throughout. This is a result of increased optical confinement and modal overlap. The peak in the shift of the dispersionless transverse mode near $600$nm is a result of competition between an increased optical confinement with decreasing diameter and a decrease in the incoupling as a result of falling optical cross section.

\subsection*{Parametric Scattering}
\begin{figure} 
\includegraphics[width=0.8\textwidth]{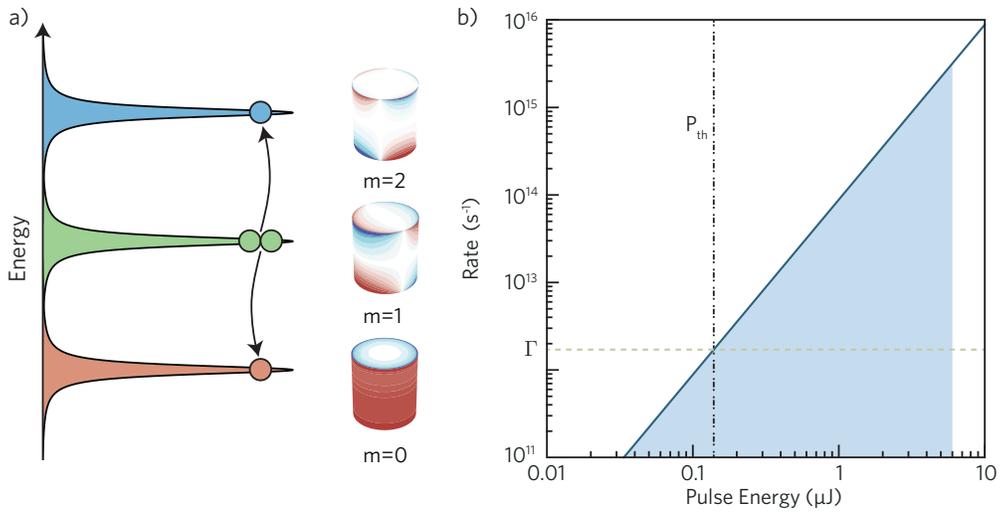}
\caption{\label{fig:Fig3} a. An illustration of parametric scattering between 3 modes with different values of the azimuthal  quantum number $m$. Illustrated are the $z-$component of the electric field for these modes in a cylinder of height $800$nm and of diameter $850$nm. b. Plot of the two sides of the threshold condition in Eq.~\ref{eq:Threshold} as a function of the pulse energy. The stimulated in-scattering is show by a solid line and the loss rate by a dashed line. They cross at the threshold power, indicated by the vertical dot-dashed line. The shading indicates the region experimentally accessible with the laser power reported by the group of A. Paarmann\cite{Passler17}.}
\end{figure}
Having shown that the $\chi^{(3)}$ nonlinearity in SiC is expected to be large enough to lead to experimentally measurable effects, we now investigate the feasibility of a polariton parametric oscillator based on phonon-polaritons. Parametric oscillators exploit optical nonlinearities to convert photons from a pump beam scattering into photons in signal and idler channels. This requires the three modes involved to respect phase matching conditions that can put strong constraints on the feasibility of practical devices. In microcavity polariton systems phase matching can be achieved at a magic angle, relying on the kink in the polariton dispersion \cite{Baumberg00}, or utilising multiple cavities \cite{Diederichs06}. In this regard localised phonon polaritons are a much more flexible system in which to study parametric oscillation due to the morphologically dependant nature of their resonances.\\
We study the general process where two excitations in the central pump mode with frequency $\omega_{C}$, containing a coherent beam with $\mathcal{N}_C\gg1$ excitations,
scatter to one in the upper (idler) and one in the lower (signal) empty modes with frequencies $\omega_{L}$ and $\omega_{U}$. Using the Fermi golden rule the spontaneous emission rate can be put in the form
\begin{align}
	\mathcal{W} =  \frac{3^4 \pi \hbar^2 \omega_{C}^2 \mathcal{N}_C^2}{2^3\epsilon_0^2 \mathrm{V}_{C}^2} \int & \mathrm{d} \omega \;  \omega \left(2 \omega_{C} - \omega \right) \rho_{L}\left(\omega \right)  \rho_{U}\left(2 \omega_{C} - \omega \right) 
	\lvert \tilde{\chi}^{(3)} \left(\omega_{C},\omega_{C},\omega,2 \omega_{C} - \omega\right)   \rvert^2,
	\label{eq:FGRp}
\end{align}
where $\mathcal{N}_C$ is the number of excitations in the pumped mode, $\rho_{\mathbf{p}}\left(\omega\right)$ represents the density of final states of mode $\mathbf{p}$, and $\tilde{\chi}^{(3)}$ is the modal third-order nonlinear coefficient as defined in the Supplementary Information. The product of two output modes densities implies non-negligible emission  only if the parametric scattering condition $\omega_{L} + \omega_{U} = 2 \omega_{C}$ is satisfied within the modes linewidths.\\
Although our theory is applicable to any resonator geometry and can easily be used for practical devices such as periodic arrays of resonators on a same material substrate where the analogy with microcavity polaritonics is most apparent \cite{Gubbin16a}, here we apply it to the illustrative system of an isolated cylindrical resonator. 
The modes of a cylinder vary azimuthally as $e^{i m \phi}$, in this case we consider two transverse $m=1$ modes scattering to one longitudinal $m=0$ mode and one $m=2$ mode as illustrated in Fig.~\ref{fig:Fig3}a. Such a combination has no azimuthal variation, except that arising from the $\chi^{(3)}$ tensor, ensuring a non-vanishing matrix element. The resonant frequencies are calculated utilising a pole-finding algorithm and the COMSOL multiphysics finite element solver and for a diameter of $850$nm we obtain $~911.5/\mathrm{cm}, \; 916.5/\mathrm{cm}, \; 921.5/\mathrm{cm}$ for the $m=1,2,3$ modes respectively with Q-factors $123,\; 133,\; 154$, fulfilling the parametric scattering condition. Following the procedure outlined in the Supplementary Information we can then write
\begin{align}
	\mathcal{W} &= \frac{3^4 \hbar^2 \omega_{C}^2 \mathcal{N}_C^2}{2^3 \epsilon_0^2 \mathrm{V}_{C}^2} \lvert \tilde{\chi}^{(3)} \left(\omega_{C},\omega_{C},\omega_{L}, \omega_{U}\right) \rvert^2  \frac{\gamma_U \gamma_L}{\left(\gamma_U + \gamma_L\right)^2}\left[ \omega_{L} \mathcal{Q}_{U} + \mathcal{Q}_{L} \omega_{U}\right],
\end{align}
where the $\mathcal{Q}_{\mathbf{p}}$ are the quality factors of the respective output modes. We can utilise this scattering rate, in conjunction with the known pump rate  and loss rates calculated from simulations of the linear electromagnetics of the resonator to form a set of rate equations linking the population of the pumped mode $\mathcal{N}_C$ to the ones of the output modes $\mathcal{N}_L$ and $\mathcal{N}_U$. Solving these equations in the undepleted regime under the assumption that the loss rates for each mode are approximately equal and given by $\gamma$, we can derive the threshold condition
\begin{equation}
	2 \mathcal{W} \mathcal{N}_{C,\mathrm{th}}^2  = \gamma,
	\label{eq:Threshold}
\end{equation}
from which we can calculate the required threshold energy per pulse utilising the free-electron lasers characteristic pulse length and spot size extrapolated to continuous wave pumping as $139\mathrm{nJ}$. We plot the two sides of this equation in Fig.~\ref{fig:Fig3}b as a function of the pulse energy, indicating the threshold by the vertical dot-dashed line and shading the region below the single pulse energy of $6 \mathrm{\mu J}$ accessible using a mid-infrared free electron laser  \cite{Passler17}.  
The threshold lies deeply inside the accessible region and this hints tantalisingly that such processes are experimentally attainable in such sub-diffraction resonator systems. Note that here we considered only one particular set of modes fulfilling phase matching conditions, chosen for sake of simplicity. Conservation relations could in fact also be achieved in different geometries, for example by coupling the modes of discrete surface phonon resonators to propagative surface phonon polaritons on a planar interface \cite{Gubbin16a}.\\

\section*{Conclusion}
In conclusion, we have presented a theory of four-wave-mixing in arbitrary phonon-polariton systems. We applied this theory to the measurement of  Kerr self-interaction \cite{DeLeon14b} for both propagative and localised  modes, showing that the relevant shifts are within experimental reach. We further studied parametric scattering between discrete resonator modes, again showing parametric oscillator regime is achievable for reasonable experimental parameters. The theory considered above can be easily applied to realistic experimental geometries where phase matching can be achieved by hybridising the modes of discrete resonators with a propagating phonon polariton \cite{Gubbin16a}. Localised phonon polaritons, thanks to their small mode volumes, long lifetimes, and large nonlinearities, present themselves as an ideal platform for mid-infrared nonlinear polaritonics. We hope the present work will stimulate further investigation in this fascinating domain, which we expect to permit translation of many of the groundbreaking results obtained with microcavity polaritons to a novel and still challenging frequency domain.
 
\section*{Acknowledgement}
The authors thank Joshua D. Caldwell for useful feedback. We acknowledge support from EPSRC grant EP/M003183/1. S.D.L. is Royal Society Research Fellow.

%%%%%%%%%% Merge with supplemental materials %%%%%%%%%%
\pagebreak
\begin{center}
\textbf{\large Supplementary Information}
\end{center}
%%%%%%%%%% Merge with supplemental materials %%%%%%%%%%
%%%%%%%%%% Prefix a "S" to all equations, figures, tables and reset the counter %%%%%%%%%%
\setcounter{equation}{0}
\setcounter{figure}{0}
\setcounter{table}{0}
\setcounter{page}{1}
\makeatletter
\renewcommand{\theequation}{S\arabic{equation}}
\renewcommand{\thefigure}{S\arabic{figure}}
\renewcommand{\bibnumfmt}[1]{[S#1]}
\renewcommand{\citenumfont}[1]{S#1}
\newcommand{\braket}[2]{\langle #1|#2\rangle}
\def\bra#1{\mathinner{\langle{#1}|}} 
\def\ket#1{\mathinner{|{#1}\rangle}} 
\def\Bra#1{\left<#1\right|} 
\def\Ket#1{\left|#1\right>}

\subsection{Derivation of the Third-Order Nonlinear Hamiltonian}
We can write the general linear Hamiltonian of a spatially inhomogeneous light-matter system in the form
\begin{align}
\label{eqn:PZWH}
\hat{\mathcal{H}}_0&=\int \mathrm{d}\mathbf{r}  \left[\frac{\hat{\mathrm{D}}^2}{2\epsilon_0}+\frac{\mu_0 \hat{\mathrm{H}}^2}{2}+\frac{\hat{\mathrm{P}}^2}{2{\rho}}
+\frac{{\rho}{\omega}_{\mathrm{LO}}^2\hat{\mathrm{X}}^2}{2}-\frac{\kappa}{\epsilon_0} \hat{\mathbf{X}}\cdot \hat{\mathbf{D}}\right],
\end{align}
where $\hat{\mathbf{D}}$ $(\hat{\mathbf{H}})$ is the electric displacement (magnetic) field operator and $\hat{\mathbf{X}}$ $(\hat{\mathbf{P}})$ is the material displacement (momentum) field operator. The system longitudinal optic phonon resonance is denoted $\omega_{\mathrm{LO}}$, it's density by $\rho$, and $\kappa$ quantifies the light-matter coupling. Such a Hamiltonian can be diagonalised by introduction of polaritonic operators, which are linear superpositions of the bare field operators
\begin{equation}
\label{eqn:Kexp}
\hat{\mathcal{K}}_{\mathbf{p}}=\int \mathrm{d}\mathbf{r} \left[\boldsymbol{\alpha}_{\mathbf{p}}(\mathbf{r})\cdot \hat{\mathbf{D}}(\mathbf{r}) +\boldsymbol{\beta}_{\mathbf{p}}(\mathbf{r})\cdot \hat{\mathbf{H}}(\mathbf{r}) +\boldsymbol{\gamma}_{\mathbf{p}}(\mathbf{r})\cdot \hat{\mathbf{P}}(\mathbf{r})+\boldsymbol{\eta}_{\mathbf{p}}(\mathbf{r})\cdot \hat{\mathbf{X}}(\mathbf{r})\right],
\end{equation}
where the Greek letters represent spatially varying vectorial Hopfield coefficients and $\mathbf{p}$ runs over both discrete and continuous parts of the spectrum. Following the approach developed in a previous publication \cite{SGubbin16b} we can write the linear Hamiltonian in diagonal form as 
\begin{equation}
	\hat{\mathcal{H}}_{0} = \sum_{\mathbf{p}} \hbar \omega_{\mathbf{p}} \hat{\mathcal{K}}_{\mathbf{p}}^{\dag} \hat{\mathcal{K}}_{\mathbf{p}}.
\end{equation}
Such an approach also allows us to account for higher order terms. A general $(N-1)^{\text{th}}$ order nonlinear term can be written in the form
\begin{align}
\label{eqn:HNL}
\mathcal{H}_{\mathrm{NL}}^{(N)}&=\int \mathrm{d}\mathbf{r}\, \sum_{t_1\cdots t_N}\sum_{j_1\cdots j_N=1}^3\Phi^{j_1\cdots j_{N}}_{t_1\cdots t_{N}}\prod_{{l}=1}^{N} \hat{\mathcal{O}}^{j_l}_{t_l},
\end{align}
where $\hat{\mathcal{O}}^{j_l}_{t_l}$ is the Cartesian component $j_l$ of operator $\hat{\boldsymbol{\mathcal{O}}}_{t_l}$
 and $\Phi^{j_1\cdots j_{N}}_{t_1\cdots t_{N}}$ the coupling tensor. 
 In the case $N=4$ of interest for this work, four terms can contribute to the nonlinear tensor: $\hat{\mathbf{X}}\hat{\mathbf{X}}\hat{\mathbf{X}}\hat{\mathbf{X}},\; \hat{\mathbf{X}}\hat{\mathbf{X}}\hat{\mathbf{X}}\hat{\mathbf{E}}, \; \hat{\mathbf{X}}\hat{\mathbf{X}}\hat{\mathbf{E}}\hat{\mathbf{E}},\; \hat{\mathbf{X}} \hat{\mathbf{E}} \hat{\mathbf{E}} \hat{\mathbf{E}}$, and $\hat{\mathbf{E}} \hat{\mathbf{E}} \hat{\mathbf{E}} \hat{\mathbf{E}}$. Exploiting the linear relationship between the lattice displacement $\hat{\mathbf{X}}$ and the electric field $\hat{\mathbf{E}}$  it is thus possible to derive the general electrical quartic nonlinear Hamiltonian \cite{SGubbin17b} as
\begin{equation}
	\hat{\mathcal{H}}^{(3)} = \frac{3}{4} \epsilon_0 \sum_{ijkl} \sum_{\left[\mathbf{p}_1-\mathbf{p}_4\right]} \int \mathrm{d}\mathbf{r} \;\;  \chi_{ijkl}^{(3)} \left(\mathbf{r}, \omega_{\mathbf{p}_1},\omega_{\mathbf{p}_2},\omega_{\mathbf{p}_3},\omega_{\mathbf{p}_4}\right) \hat{\mathrm{E}}_{\mathbf{p}_1}^{i}\left(\mathbf{r}\right) \hat{\mathrm{E}}_{\mathbf{p}_2}^{j}\left(\mathbf{r}\right)\hat{\mathrm{E}}_{\mathbf{p}_3}^{k}\left(\mathbf{r}\right)\hat{\mathrm{E}}_{\mathbf{p}_4}^{l}\left(\mathbf{r}\right). \label{eq:HamNL}
\end{equation}
The general form of the third order nonlinear susceptibility in the bulk is given by
\begin{align}
	\chi^{(3)} \left(\omega_{1},\omega_{2},\omega_{3},\omega_{4}\right)  = \chi_{\infty}^{(3)} \biggr[& 1 + C_1 \frac{1}{D\left(\omega_{1}\right) D\left(\omega_{2}\right) D\left(\omega_{3}\right) D\left(\omega_{4}\right)} \\ & + C_2\left(\frac{1}{D\left(\omega_{1}\right) D\left(\omega_{2}\right) D\left(\omega_{3}\right)} + \frac{1}{D\left(\omega_{1}\right) D\left(\omega_{2}\right) D\left(\omega_{4}\right)} +  \frac{1}{D\left(\omega_{2}\right) D\left(\omega_{3}\right) D\left(\omega_{4}\right)} \right) \nonumber\\ & + C_3\left(\frac{1}{D\left(\omega_{1}\right) D\left(\omega_{2}\right) } + \frac{1}{D\left(\omega_{1}\right) D\left(\omega_{3}\right)} +  \frac{1}{D\left(\omega_{1}\right) D\left(\omega_{4}\right)} +  \frac{1}{D\left(\omega_{2}\right) D\left(\omega_{3}\right)} \right. \nonumber\\ & \quad \quad \quad \left. +  \frac{1}{D\left(\omega_{2}\right) D\left(\omega_{4}\right)} +  \frac{1}{D\left(\omega_{3}\right) D\left(\omega_{4}\right)}\right) \nonumber\\ & + C_4 \left(\frac{1}{D\left(\omega_{1}\right)} + \frac{1}{ D\left(\omega_{2}\right) }+\frac{1}{D\left(\omega_{3}\right)} + \frac{1}{D\left(\omega_{4}\right)} \right) \biggr],\nonumber
\end{align}
where $D\left(\omega\right) = 1 - \frac{\omega^2}{\omega_{\mathrm{TO}}^2}$, $\omega_{\mathrm{TO}}$ is the crystal's transverse optical frequency and we suppressed the cartesian indexes for brevity. The high-frequency, static, contribution to the nonlinearity arising from the $\mathbf{\hat{E}\hat{E}\hat{E}\hat{E}}$ is given by $\chi_{\infty}^{(3)}$ and the $C_{i}$ describe the contributions of the different terms.\\

The term proportional to $C_{1}$ is the pure mechanical anharmonicity arising from phonon-phonon scattering and is expected to dominate when all participating frequencies are in the neighbourhood of $\omega_{\mathrm{TO}}$, due to the fourth order pole proportional to $1/D\left(\omega\right)^4$. Being purely mechanical in nature, this term can be probed by pure lattice dynamics calculations such as those carried out by Vanderbilt {\it et al.} for a variety of crystals with diamond structure \cite{SVanderbilt86}. Previously it was shown that the material coefficients derived in this way can describe with acceptable accuracy the second-order nonlinear susceptibility of $\beta$-SiC. Here we assume that this universality also holds to third-order, calculating for the three non-zero components of the screened $\chi^{(3)}$ non-linear susceptibility
 \begin{align}
 	\left[\chi_{\infty}^{(3)} C_1 \right]_{xxxx} &= - 0.46 \times 10^{-20} \frac{\mathrm{m^2}}{\mathrm{V}^2},\\
 	\left[\chi_{\infty}^{(3)} C_1 \right]_{xxyy} &= - 0.84 \times 10^{-20} \frac{\mathrm{m^2}}{\mathrm{V}^2},\nonumber\\
 	\left[\chi_{\infty}^{(3)} C_1 \right]_{xyxy} &= - 0.21 \times 10^{-20} \frac{\mathrm{m^2}}{\mathrm{V}^2}.	\nonumber
 \end{align}
 
 \subsection{Derivation of the Kerr Self-Interaction}
We can write the electric field in terms of the polaritonic operator $\hat{\mathcal{K}}_{\mathbf{p}}$
\begin{equation}
	\hat{\mathbf{E}}_{\mathbf{p}}\left(\mathbf{r}\right) = \sum_{\mathbf{p}} \sqrt{\frac{\hbar \omega_{\mathbf{p}}}{2 \epsilon_0 \mathrm{V}_{\mathbf{p}}}} \left(\bar{\boldsymbol{\alpha}}_{\mathbf{p}}\left(\mathbf{r}\right) \hat{\mathcal{K}}_{\mathbf{p}} + \boldsymbol{\alpha}_{\mathbf{p}} \left(\mathbf{r}\right)\hat{\mathcal{K}}_{\mathbf{p}}^{\dag} \right),
\end{equation} 
where $\mathrm{V}_{\mathbf{p}}$ is the photonic mode volume and the Hopfield coefficients $\boldsymbol{\alpha}_{\mathbf{p}}$ are normalised so
\begin{equation}
	\left[\bar{\boldsymbol{\alpha}}_{\mathbf{p}} \cdot \boldsymbol{\alpha}_{\mathbf{p}}\right]_{\mathrm{max}} = 1.
\end{equation}
Substituting this into the nonlinear Hamiltonian yields 
\begin{multline}
\label{eqn:HamNLm}
	\hat{\mathcal{H}}^{(3)} = \frac{9}{8} \frac{\hbar^2}{\epsilon_0}   \sum_{\left[\mathbf{p}_1-\mathbf{p}_4\right]} \sqrt{\frac{\omega_{\mathbf{p}_1} \omega_{\mathbf{p}_2} \omega_{\mathbf{p}_3} \omega_{\mathbf{p}_4}}{\mathrm{V}_{\mathbf{p}_1} \mathrm{V}_{\mathbf{p}_2} \mathrm{V}_{\mathbf{p}_3} \mathrm{V}_{\mathbf{p}_4}}}\\ \times \sum_{ijkl}  \int \mathrm{d}\mathbf{r} \;  \chi_{ijkl}^{(3)} \left(\mathbf{r}, \omega_{\mathbf{p}_1},\omega_{\mathbf{p}_2},\omega_{\mathbf{p}_3},\omega_{\mathbf{p}_4}\right)   \alpha_{\mathbf{p}_1}^{i}\left(\mathbf{r}\right) \alpha_{\mathbf{p}_2}^{j}\left(\mathbf{r}\right) \bar{\alpha}_{\mathbf{p}_3}^k\left(\mathbf{r}\right) \bar{\alpha}_{\mathbf{p}_4}^l\left(\mathbf{r}\right) \hat{\mathcal{K}}_{\mathbf{p}_1}^{\dag} \hat{\mathcal{K}}_{\mathbf{p}_2}^{\dag} \hat{\mathcal{K}}_{\mathbf{p}_3} \hat{\mathcal{K}}_{\mathbf{p}_4},
\end{multline}
and specialising onto the case where $\mathbf{p}_1 = \mathbf{p}_2 = \mathbf{p}_3 = \mathbf{p}_4 = \mathbf{p}$ we can write the Kerr self-interction Hamiltonian in the form
\begin{equation}
	\hat{\mathcal{H}}^{(3)}_{0} =   \frac{9}{8}  \frac{\hbar^2 \omega_{\mathbf{p}}^2}{\epsilon_0 \mathrm{V}_{\mathbf{p}}^2}  \sum_{ijkl}  \int \mathrm{d}\mathbf{r} \;  \chi_{ijkl}^{(3)} \left(\mathbf{r}, \omega_{\mathbf{p}},\omega_{\mathbf{p}},\omega_{\mathbf{p}},\omega_{\mathbf{p}}\right)   \bar{\alpha}_{\mathbf{p}}^{i}\left(\mathbf{r}\right) \bar{\alpha}_{\mathbf{p}}^{j}\left(\mathbf{r}\right) \alpha_{\mathbf{p}}^k\left(\mathbf{r}\right) \alpha_{\mathbf{p}}^l\left(\mathbf{r}\right) \hat{\mathcal{K}}_{\mathbf{p}}^{\dag} \hat{\mathcal{K}}_{\mathbf{p}}^{\dag} \hat{\mathcal{K}}_{\mathbf{p}} \hat{\mathcal{K}}_{\mathbf{p}}.
\end{equation}
We can at this point linearise this Hamiltonian. Assuming a large population of the mode we can expand the operator $\hat{\mathcal{K}}_{\mathbf{p}}$ around the mean occupation number $\mathcal{N}_{\mathbf{p}}$ as $\hat{\mathcal{K}}_{\mathbf{p}} = \sqrt{\mathcal{N}_{\mathbf{p}}} + \hat{\delta \mathcal{N}}_{\mathbf{p}}$. This yields for the resonant, quadratic component 
\begin{equation}
	\hat{\Delta \mathcal{H}}^{(3)}_{0} =   \frac{9}{2}  \frac{\hbar^2 \omega_{\mathbf{p}}^2 \mathcal{N}_{\mathbf{p}}}{\epsilon_0 \mathrm{V}_{\mathbf{p}}^2}  \sum_{ijkl}  \int \mathrm{d}\mathbf{r} \;  \chi_{ijkl}^{(3)} \left(\mathbf{r}, \omega_{\mathbf{p}},\omega_{\mathbf{p}},\omega_{\mathbf{p}},\omega_{\mathbf{p}}\right)   \bar{\alpha}_{\mathbf{p}}^{i}\left(\mathbf{r}\right) \bar{\alpha}_{\mathbf{p}}^{j}\left(\mathbf{r}\right) \alpha_{\mathbf{p}}^k\left(\mathbf{r}\right) \alpha_{\mathbf{p}}^l\left(\mathbf{r}\right) \hat{\delta \mathcal{N}}_{\mathbf{p}}^{\dag} \hat{\delta \mathcal{N}}_{\mathbf{p}},
\end{equation}
and thus the fractional frequency shift is given by the expression
\begin{equation}
	\Delta_{\mathbf{p}} =  \frac{9}{2} \frac{\hbar \omega_{\mathbf{p}} \mathcal{N}_{\mathbf{p}}}{\epsilon_0 \mathrm{V}_{\mathbf{p}}^2}  \sum_{ijkl}  \int \mathrm{d}\mathbf{r} \;  \chi_{ijkl}^{(3)} \left(\mathbf{r}, \omega_{\mathbf{p}},\omega_{\mathbf{p}},\omega_{\mathbf{p}},\omega_{\mathbf{p}}\right)   \bar{\alpha}_{\mathbf{p}}^{i}\left(\mathbf{r}\right) \bar{\alpha}_{\mathbf{p}}^{j}\left(\mathbf{r}\right) \alpha_{\mathbf{p}}^k\left(\mathbf{r}\right) \alpha_{\mathbf{p}}^l\left(\mathbf{r}\right).
\end{equation}

\subsection{Derivation of the Parametric Scattering Rate}
Assuming that two excitations from the same pump mode with frequency $\omega_{{{C}}}$ are annihilated, we can calculate the scattering rate perturbatively via the Fermi golden rule as
\begin{equation}
	\mathcal{W} = \frac{2 \pi}{\hbar^2} \sum_{{\mathbf{p}_{1},\mathbf{p}_{1}}}  \delta\left(2  \omega_{C} - \omega_{\mathbf{p}_1} - \omega_{\mathbf{p}_2}\right)  \lvert  \langle f_{\mathbf{p}_1,\mathbf{p}_2} \lvert \mathcal{H}^{(3)} \rvert i \rangle  \rvert^2,
\end{equation}
where $\ket{i}$ is the system initial state and $\ket{f_{\mathbf{p}_1,\mathbf{p}_2}}$ the final state with excitations in channels labelled $\mathbf{p}_1, \mathbf{p}_2$. Assuming the initial state is a coherent state with $\mathcal{N}_{{C}}\gg1$ excitations in the input mode and none in the two output branches, this yields the spontaneous scattering rate
\begin{equation}
	\mathcal{W} = \frac{3^4 \pi \mathcal{N}_{C}^2}{2^3\hbar^2} \frac{\hbar^4 \omega_{{C}}^2}{\epsilon_0^2 \mathrm{V}_{{C}}^2}\sum_{{\mathbf{p}_{1},\mathbf{p}_{2}}} \delta\left(2  \omega_{{C}} - \omega_{\mathbf{p}_1} - \omega_{\mathbf{p}_2}\right) \omega_{\mathbf{p}_{1}}  \omega_{\mathbf{p}_{2}}  \lvert\tilde{\chi}^{(3)}\left(\omega_{{C}},\omega_{{C}},\omega_{\mathbf{p}_{1}},\omega_{\mathbf{p}_{2}}\right) \rvert^2,
\end{equation}
where the modal nonlinear coefficient renormalised by the interaction volume is given by
\begin{equation}
\label{Overlap}
\tilde{\chi}^{(3)}\left(\omega_{{C}},\omega_{{C}},\omega_{\mathbf{p}_{1}},\omega_{\mathbf{p}_{2}}\right)=\sum_{ijkl} 
\int \mathrm{d}\mathbf{r} \;\; \frac{\chi^{(3)}_{ijkl}\left(\mathbf{r},\omega_{{C}},\omega_{{C}},\omega_{\mathbf{p}_{1}},\omega_{\mathbf{p}_{2}}\right)}{\sqrt{\mathrm{V}_{\mathbf{p}_{1}} \mathrm{V}_{\mathbf{p}_{2}}}}  \alpha_{C}^{i} \left(\mathbf{r}\right) \alpha_{C}^{j} \left(\mathbf{r}\right) \bar{\alpha}_{\mathbf{p}_1}^{k} \left(\mathbf{r}\right)\bar{\alpha}_{\mathbf{p}_2}^{l} \left(\mathbf{r}\right).
\end{equation}
We introduce at this point the densities of states $\rho_{L}(\omega)$ and $\rho_{U}(\omega)$ over the lower and upper branches, which we supposed to be Lorentzian with central frequencies $\omega_{L}$ and $\omega_{U}$ and FWHM broadenings $\gamma_{L}$ and $\gamma_{U}$.
 Using those densities we can convert the sums over the output frequencies to integrals, 
\begin{align}
	\mathcal{W} &= \frac{3^4 \pi \mathcal{N}_{C}^2 \hbar^2 \omega_{{C}}^2}{2^3 \epsilon_0^2 \mathrm{V}_{{C}}^2}  \nonumber \int \mathrm{d} \omega_{\mathbf{p}_1} \rho_{L}\left(\omega_{\mathbf{p}_1}\right)   \int \mathrm{d} \omega_{\mathbf{p}_2} \rho_{U}\left(\omega_{\mathbf{p}_2}\right)  \delta\left(2  \omega_{{C}} - \omega_{\mathbf{p}_1} - \omega_{\mathbf{p}_2}\right) \omega_{\mathbf{p}_{1}}  \omega_{\mathbf{p}_{2}}   \lvert \tilde{\chi}^{(3)}\left(\omega_{{C}},\omega_{{C}},\omega_{\mathbf{p}_{1}},\omega_{\mathbf{p}_{2}}\right) \rvert^2,\\
	&=\frac{3^4 \pi \mathcal{N}_{C}^2 \hbar^2 \omega_{{C}}^2}{2^3 \epsilon_0^2 \mathrm{V}_{{C}}^2} \int \mathrm{d} \omega_{\mathbf{p}_1} \rho_{L}\left(\omega_{\mathbf{p}_1}\right)   \rho_{U}\left(2 \omega_{C} - \omega_{\mathbf{p}_1} \right)  \omega_{\mathbf{p}_{1}}  \left(2 \omega_{C} - \omega_{\mathbf{p}_1} \right)  \lvert \tilde{\chi}^{(3)}\left(\omega_{{C}},\omega_{{C}},\omega_{\mathbf{p}_{1}},2\omega_{{C}}-\omega_{\mathbf{p}_{1}}\right) \rvert^2.\nonumber
\end{align}
The integrals can be carried out by integration in the complex plane, noting that the central frequencies of the Lorentzian's relate to each other by $2 \omega_{{C}} = \omega_{L} + \omega_{U}$. Assuming that the $\chi^{(3)}$ is constant over the resonator linewidths we can solely consider the dispersive terms inside the integrand and outside of the modulus. The integral is then given by 
\begin{align}
	2 \pi i \sum \mathrm{Res}\left[ \rho_{L}\left(\omega\right)   \rho_{U}\left(2 \omega_{C} - \omega \right)  \omega  \left(2 \omega_{C} - \omega\right) \right] &= \frac{\gamma_{L} \gamma_{U}}{\pi\left(\gamma_{L} + \gamma_{U}\right)^2} \left[ \omega_{U} \mathcal{Q}_{{L}} + \mathcal{Q}_{U} \omega_{{L}}\right],
\end{align}
where $ \sum \mathrm{Res}\left[g(\omega)\right]$ is the sum over the residues of $g(\omega)$, we assumed narrow linewidths $\omega_{L}, \omega_{U} \gg \gamma_{L}, \gamma_{U}$, and we introduced the modal quality factors of the $\mathbf{p}$ mode $\mathcal{Q}_{\mathbf{p}} = \omega_{\mathbf{p}}/\gamma_{\mathbf{p}}$. This yields the final scattering rate
\begin{align}
	\mathcal{W} 	&=\frac{3^4 \mathcal{N}_{C}^2 \hbar^2 \omega_{{C}}^2}{2^3 \epsilon_0^2 \mathrm{V}_{{C}}^2}    \lvert \tilde{\chi}^{(3)}\left(\omega_{{C}},\omega_{{C}},\omega_{L},  \omega_{U} \right)
	\rvert^2 \frac{\gamma_{L} \gamma_{U}}{\left(\gamma_{L} + \gamma_{U}\right)^2} \left[ \omega_{U} \mathcal{Q}_{L} + \mathcal{Q}_{U} \omega_{L}\right].
	\label{eq:Wfinal}
\end{align}

We can at this point utilise the scattering rate in Eq.~\ref{eq:Wfinal} to write a rate equation for modes $\mathcal{N}_{{L}},\mathcal{N}_{{U}}\ll 1$ 
\begin{align}
\label{eq:Rates}
	\dot{\mathcal{N}}_{{L}} &=  - \gamma_{{L}} \mathcal{N}_{{L}} + \mathcal{W} (\mathcal{N}_{{L}}+1)( \mathcal{N}_{{U}}+1) \mathcal{N}_{C}^2,\\
	\dot{\mathcal{N}}_{{U}} &=  - \gamma_{{U}} \mathcal{N}_{{U}} +\mathcal{W} (\mathcal{N}_{{L}}+1)( \mathcal{N}_{{U}}+1)  \mathcal{N}_{C}^2.\nonumber
	\end{align}
Assuming all loss rates to be equal $\gamma = \gamma_{{U}} = \gamma_{{L}}$ we obtain 
$\mathcal{N}_{{L}}=\mathcal{N}_{{U}}$, simplify Eq.~\ref{eq:Rates} to the single equation 
\begin{align}
	\dot{\mathcal{N}}_{{L}} &=  - \gamma_{{L}} \mathcal{N}_{{L}} + \mathcal{W} (\mathcal{N}_{{L}}+1)^2\mathcal{N}_{C}^2\simeq (- \gamma_{{L}}  + 2\mathcal{W} \mathcal{N}_{C}^2) \mathcal{N}_{{L}} +\mathcal{W} \mathcal{N}_{C}^2 ,
	\end{align}
from which we can directly read the threshold condition
\begin{equation}
	\mathcal{N}_{C}^{\mathrm{th}} = \sqrt{\frac{\gamma}{2 \mathcal{W}}}.
\end{equation}


\begin{thebibliography}{}

\bibitem{Houdre94} Houdr{\'e}, R.; Stanley, R. P.; Oesterle, U.; Ilegems, M.; Weisbuch, C. Room-temperature cavity polaritons in a semiconductor microcavity. {\it Phys. Rev. B} 1994, {\bf 49}, 16761.

\bibitem{Kasprzak06} Kasprzak, J.; Richard, M.; Kundermann, S.; Baas, A.; Jeambrun, P.; Keeling, J. J. M.; Marchett, F. M.; Szyma{\'n}ska, M. H.; Andr{\'e}, R.; Staehli, J. L.; Savona, V.; Littlewood, P. B.; Deveaud, B.; Dang, L. S. Bose Einstein condensation of exciton polaritons. {\it Nature} 2006, {\bf 443}, 409-414.

\bibitem{Amo09} Amo, A.; Sanvitto, D.; Laussy, F. P.; Ballarini, D.; del Valle, E.; Martin, M. D.; Lema{\'i}tre, A.; Bloch, J.; Krizhanovskii, D. N.; Skolnick, M. S.; Tejedor, C.; Vi{\~n}a, L. Collective fluid dynamics of a polariton condensate in a semiconductor microcavity. {\it Nature} 2009, {\bf 457}, 291-295.

\bibitem{Baumberg00} Baumberg, J. J.; Savvidis, P. G.; Stevenson, R. M.; Tartakovskii, A. I.; Skolnick, M. S.; Whittaker, D. M.; Roberts, J. S. Parametric oscillation in a vertical microcavity: A polariton condensate or micro-optical parametric oscillation. {\it Phys. Rev. B} 2000, {\bf 62}, R16247.

\bibitem{Diederichs06} Diederichs, C.; Tignon, J.; Dasbach, G.; Ciuti, C.; Lema{\'i}tre, A.; Bloch, J.; Roussignol, P.; Delalande, C. Parametric oscillation in vertical triple microcavities. {\it Nature} 2006, {\bf 440}, 904-907.

\bibitem{Bhattacharya13} Bhattacharya, P.; Xiao, B.; Das, A.; Bhowmick, S.; Heo, J. Solid state electrically injected exciton-polariton laser. {\it Phys. Rev. Lett.} 2013, {\bf 110}, 206403.

\bibitem{Khurgin15} Khurgin, J. How to deal with the loss in plasmonics and metamaterials. {\it Nature Nanotechnology} 2015, {\bf 10}, 2-6.

\bibitem{Caldwell15} Caldwell, J. D.;  Lindsay, L.; Giannini, V.; Vurgaftman, I.; Reinecke, T. L; Maier, S. A.; Glembocki, O. J. Low-loss, infrared and terahertz nanophotonics using surface phonon polaritons. {\it Nanophotonics} 2015, {\bf 4}, 44-68.

\bibitem{Schuller09} Schuller, J. A.; Taubner, T.; Brongersma, M. L.; Optical antenna thermal emitters. {\it Nature Photonics} 2009. {\bf 3}, 658-661.

\bibitem{Gubbin17a} Gubbin, C. R.; Maier, S. A.; De Liberato, S.; Theoretical Investigation of Phonon Polaritons in SiC Micropillar Resonators. {\it Phys. Rev. B} 2017, {\bf 95}, 035313.

\bibitem{Wang17} Wang, T.; Li, P.; Chigrin, D. N.; Giles, A. J.; Bezares, F. J.; Glembocki, O. J.; Caldwell, J. D.; Taubner, T. Phonon-Polaritonic Bowtie Nanoantennas: Controlling Infrared Thermal Radiation at the Nanoscale. {\it ACS Photonics} 2017, {\bf 4}, 1753-1760.

\bibitem{Caldwell13} Caldwell, J. D.; Glembocki, O. J.; Francescato, Y.; Sharac, N.; Giannini, V.; Bezares, F. J.; Long, J. P.; Owrutsky, J. C.; Vurgaftman, I.; Tischler, J. G.; Wheeler, V. G.; Bassim, N. B.; Shirey, L. M.; Kasica, R.; Maier, S. A. Low-Loss, Extreme Subdiffraction Photon Confinement via Silicon Carbide Localized Surface Phonon Polariton Microresonators. {\it Nano Lett.} 2013, {\bf 13}, 3690-3697.

\bibitem{Chen14} Chen, Y.; Francescato, Y.; Caldwell, J. D.; Giannini, V.; Ma{\ss}, T. W. W.; Glembocki, O. J.; Bezares, F. J. Taubner, T.; Kasica, R.; Hong, M.; Maier, S. A. Spectral Tuning of Localized Surface Phonon Polariton Resonators for Low-loss Mid-IR Applications. {\it ACS Phot.} 2014, {\bf 1,} 718-724.
	
\bibitem{Gubbin16b} Gubbin, C. R.; Maier, S. A.; De Liberato, S. Real-Space Hopfield Diagonalization of Inhomogeneous Dispersive Media. {\it Phys. Rev. B} 2016, {\bf 94}, 205301.

\bibitem{Gubbin17b} Gubbin, C. R.; De Liberato, S.;  Theory of nonlinear polaritonics: $\chi^{(2)}$ scattering on a $\beta$-SiC surface. {\it ACS Photonics} 2017, {\bf 4}, 1381-1388.

\bibitem{Paarmann16} Paarmann, A.; Razdolski, I.; Gewinner, S.; Sch{\"o}llkopf, W.; Wolf, M. Effects of Crystal Anisotropy on Optical Phonon Resonances in Midinfrared Second Harmonic Response of SiC. {\it Phys. Rev. B} 2016, {\bf 94}, 134312.

\bibitem{Passler17}  Passler, N. C.; Razdolski, I.; Gewinner, S.; Sch{\"o}llkopf, W.; Wolf, M.; Paarmann, A. Second-Harmonic Generation from Critically Coupled Surface Phonon Polaritons. {\it ACS Photonics} 2017, {\bf 4}, 1048-1053.

\bibitem{Razdolski16} Razdolski, I.; Chen, Y.; Giles, A. J.; Gewinner, S.; Sch{\"o}llkopf, W.; Hong, M.; Wolf, M.; Giannini, V.; Caldwell, J. D.; Maier, S. A.; Paarmann, A.; Resonant Enhancement of Second-Harmonic Generation in the Mid-Infrared Using Localized Surface Phonon Polaritons in Subdiffractional Nanostructures. {\it Nano Lett.} 2016, {\bf 16}, 6954-6959.

\bibitem{DeLeon14} De Leon, I.; Sipe, J. E.; Boyd, R. W. Self-phase-modulation of surface plasmon polaritons. {\it Phys. Rev. A} 2014, {\bf 89}, 013855.

\bibitem{Vanderbilt86} Vanderbilt, D.; Louie, S. G.; Cohen, M. L. Calculation of Anharmonic Phonon Couplings in C, Si and Ge. {\it Phys. Rev. B} 1986, {\bf 33}, 8740-8747.

\bibitem{DeLeon14b} De Leon, I.; Shi, Z.; Liapis, A. C.; Boyd, R. W. Measurement of the complex nonlinear optical response of a surface plasmon-polariton. {\it Opt. Lett.} 2014, {\bf 39}, 2274-2277.

\bibitem{Calusine14} Calusine, G.; Politi, A.; Awschalom, D.; Silicon carbide photonic crystal cavities with integrated color centers. {\it App. Phys. Lett.} 2014, {\bf 105}, 011123.

\bibitem{Gubbin16a} Gubbin, C. R.; Martini, F.; Politi, A.; Maier, S. A.; De Liberato, S. Strong and Coherent Coupling Between Localised and Propagating Phonon Polaritons. {\it Phys. Rev. Lett.} 2016, {\bf 116}, 246402.

\end{thebibliography}

\begin{thebibliography}{11}

\bibitem{SGubbin16b} Gubbin, C. R.; Maier, S. A.; De Liberato, S. Real-Space Hopfield Diagonalization of Inhomogeneous Dispersive Media. {\it Phys. Rev. B} 2016, {\bf 94}, 205301.

\bibitem{SGubbin17b} Gubbin, C. R.; De Liberato, S.;  Theory of nonlinear polaritonics: $\chi^{(2)}$ scattering on a $\beta$-SiC surface. {\it ACS Photonics} 2017, {\bf 4}, 1381-1388.

\bibitem{SVanderbilt86} Vanderbilt, D.; Louie, S. G.; Cohen, M. L. Calculation of Anharmonic Phonon Couplings in C, Si and Ge. {\it Phys. Rev. B} 1986, {\bf 33}, 8740-8747.

\end{thebibliography}
\end{document}